# On the anisotropy barrier reduction in fast relaxing $Mn_{12}$ single-molecule magnets

Stephen Hill,[1] Muralee Murugesu[2,†] and George Christou[2]

[1]National High Magnetic Field Laboratory and Department of Physics, Florida State University, Tallahassee, Florida 32310

[2]Department of Chemistry, University of Florida, Gainesville, Florida 32611

**Abstract**

A novel angle-swept high-frequency electron paramagnetic resonance (HFEPR) technique is described that facilitates efficient *in-situ* alignment of single-crystal samples containing low-symmetry magnetic species such as single-molecule magnets (SMMs). This cavity-based technique involves recording HFEPR spectra at fixed frequency and field, while sweeping the applied field orientation. The method is applied to the study of a low-symmetry Jahn-Teller variant of the extensively studied spin $S = 10$ $Mn_{12}$ SMMs (e.g. $Mn_{12}$-acetate). The low-symmetry complex also exhibits SMM behavior, but with a significantly reduced effective barrier to magnetization reversal ($U_{eff} \approx 43$ K) and, hence, faster relaxation at low temperature in comparison with the higher-symmetry species. $Mn_{12}$ complexes that crystallize in lower symmetry structures exhibit a tendency for one or more of the Jahn-Teller axes associated with the $Mn^{III}$ atoms to be abnormally oriented, which is believed to be the cause of the faster relaxation. An extensive multi-high-frequency angle- and field-swept electron paramagnetic resonance study of $[Mn_{12}O_{12}(O_2CCH_2Bu^t)_{16}(H_2O)_4]\cdot CH_2Cl_2\cdot MeNO_2$ is presented in order to examine the influence of the abnormally oriented Jahn-Teller axis on the effective barrier to magnetization reversal. The reduction in the axial anisotropy, $D$, is found to be insufficient to account for the nearly 40% reduction in $U_{eff}$. However, the reduced symmetry of the $Mn_{12}$ core gives rise to a very significant 2nd order transverse (rhombic) zero-field-splitting anisotropy, $E \approx D/6$. This, in turn, causes a significant mixing of spin projection states well below the top of the classical anisotropy barrier. Thus, magnetic quantum tunneling is the dominant factor contributing to the effective barrier reduction in fast relaxing $Mn_{12}$ SMMs.

## I. Introduction

During the last two decades, the interdisciplinary field of molecular magnetism has evolved from the design of traditional 3D magnets composed of molecular building blocks, toward the development of molecular nanomagnets that could one day function as magnetic information storage units or qubits.[1-3] Of particular recent interest has been the synthesis and investigation of polynuclear transition metal complexes, so-called Single-Molecule Magnets (SMMs),[4] which have been shown to exhibit a range of fascinating quantum effects, including quantum magnetization tunneling (QMT)[5,6] and magnetic quantum phase interference.[7,8]

Around the same time as the first clear-cut observations of resonant QMT[5,6] in [$Mn_{12}O_{12}(O_2CCH_3)_{16}(H_2O)_4$]·$2CH_3CO_2H$·$4H_2O$ ($Mn_{12}$-acetate, or $Mn_{12}$Ac for short),[9] it was found that typical single-crystal samples contain a significant fraction (up to 10%) of fast relaxing (FR) species,[10] i.e. a minority of molecules that relax much faster than the majority slow relaxing (SR) species at a given temperature. These FR species may be identified from magnetic measurements in a variety of ways. First of all, they exhibit a peak in the out-of-phase ac susceptibility, $\chi''$, in the 2–3 K range (at 1 kHz),[10-13] as opposed to the dominant peak in the 6–7 K range associated with the majority SR species, i.e. the blocking temperature, $T_B$, associated with the FR species is considerably lower ($T_{B,FR}$ < 2 K) than for the SR molecules ($T_{B,SR}$ ~ 4 K). Consequently, the two species are sometimes referred to as 'low-temperature' (LT) and 'high-temperature' (HT) forms; henceforth, we shall mainly use the FR and SR terminology. One can also identify the FR species from very low temperature hysteresis measurements[14-16] due to the fact that they exhibit temperature-independent (pure) resonant QMT at considerably lower fields (1–2 T for T < 300 mK) than the SR molecules (3–4 T). Interestingly, well defined, evenly spaced QMT steps have been reported,[14-16] suggesting that the FR molecules and their surrounding environments are also relatively monodisperse, in spite of the fact that they are believed to be randomly distributed throughout the crystal. Furthermore, the spacing between the QMT steps associated with the FR species (~0.4 T) is not all that dissimilar to those of the SR species (~0.45 T), suggesting that they possess similar uniaxial anisotropy constants $D$.[14-16]

Subsequently, detailed synthetic work by Christou, Hendrickson and co-workers, involving ligand substitution and crystallization from a variety of solvents, resulted in the discovery of many different forms of $Mn_{12}$ possessing essentially the same neutral $Mn_{12}O_{12}$ core.[11,12,17-27] However, important differences have been inferred from X-ray and magnetic studies. Most notably, the various $Mn_{12}$ complexes can be grouped broadly into two categories, i.e. FR or SR. Indeed, a histogram of the effective magnetization relaxation barriers, $U_{eff}$ (deduced from frequency-dependent ac susceptibility measurements), for around 20 different $Mn_{12}$ complexes reveals a clear bimodal distribution,[28] with values in the 25-45 K range for the FR species and the 60-80 K range for the SR species. As we will outline below, one can essentially rule out the likelihood that this roughly factor-of-two reduction in $U_{eff}$ is due solely to a reduction in the molecular $D$, hence the motivation for the present investigation.

The synthesis of pure crystals containing neutral FR $Mn_{12}O_{12}$ species have enabled detailed characterizations of their structures and magnetic properties [we note that reduced forms of $Mn_{12}$ have also been obtained that also exhibit fast relaxation,[27-32] but these are not considered in this study]. In all cases where the FR species are obtained in single-crystal form, X-ray studies show that one or more of the Jahn-Teller (JT) elongation axes associated with the $Mn^{III}$ atoms are abnormally oriented[18,19,21,24,25,33] in comparison to the usual SR form of $Mn_{12}$.[26,27,34-37] This is illustrated in Fig. 1 for the $[Mn_{12}O_{12}(O_2CCH_2Bu^t)_{16}(H_2O)_4] \cdot CH_2Cl_2 \cdot MeNO_2$ complex[30] (hereon FR-$Mn_{12}$tBuAc), where one of the JT elongation axes has flipped (as indicated in the figure) so that it is close to parallel to the plane of the molecule, in contrast to the other seven JT axes which are approximately orthogonal to the plane of the molecule. Other $Mn_{12}$ complexes have been obtained with two and even four flipped JT axes;[21,38] these complexes also exhibit fast magnetization relaxation. It is, therefore, widely believed that this JT isomerism is responsible for the two forms of $Mn_{12}$. However, it is as yet unclear how this isomerism gives rise to the significantly different relaxation dynamics associated with the FR and SR forms.

Remarkably, it has been found that some $Mn_{12}$ complexes convert from the FR form to the SR one when left in a dry atmosphere for several weeks.[25] This conversion has been monitored for FR-$Mn_{12}$tBuAc via ac susceptibility measurements performed on a single crystal at regular time intervals.[25] Measurements performed on the fresh (wet) crystals removed directly from the mother liquor reveal a single, sharp LT $\chi''$ peak at ~2 K. As the sample ages, this peak diminishes in intensity, but does not shift. Meanwhile, a broader HT $\chi''$ peak appears at ~7 K. This behavior signifies the slow conversion between the two forms, which is attributed to the loss of volatile $CH_2Cl_2$/$MeNO_2$ solvents from the structure (also confirmed via elemental analysis). Of course, drying the samples under vacuum speeds up the transformation dramatically. Evidently, subtle differences in crystal packing forces on the $Mn_{12}$ molecules upon varying the identity of solvent molecules trapped in the crystal are sufficient to stabilize the abnormal orientation of one of the JT axes. Upon loss of solvent, the structure presumably relaxes to the normal one with more-or-less parallel JT axes on all of the eight $Mn^{III}$ atoms. Unfortunately, the dry samples do not diffract, i.e. they are disordered, in spite of the fact that the magnetic studies reveal a single, albeit broad peak in $\chi''$. Subsequent recrystallization of dry FR-$Mn_{12}$tBuAc from a $CH_2Cl_2$/MeCN mixture gives good crystals of the pure SR form (hereon SR-$Mn_{12}$tBuAc) and X-ray studies indicate that the JT axes are now in their normal orientations.[25]

It should be emphasized that there are relatively few examples of truly high-symmetry $Mn_{12}$,[26,34-37] the acetate[9] being the best known. However, when crystallized with acetic acid solvent, the resulting samples are rather disordered: not only are FR molecules randomly distributed throughout the crystal, but the acetic acid is also disordered and this is known to influence the SR species as well; this point has been discussed at length by several authors.[39-43] It could well be that it is this intrinsic disorder that also stabilizes the minority JT isomers. Interestingly, other high-symmetry forms that do not suffer from this solvent disorder also do not seem to contain a significant fraction of the FR species. The cleanest examples include: $Mn_{12}$BrAc,[26,44] $Mn_{12}$tBuAc,[36,45,46] and $Mn_{12}$Ac·MeOH,[34] i.e. the acetate crystallized from MeOH solvent instead of acetic acid and water. All of these truly high-symmetry complexes belong to the SR category. Nevertheless, high-symmetry structures do not seem absolutely necessary in order to obtain SR crystals, i.e. plenty of SR examples exist that possess low symmetry structures.[35] It seems that a very significant structural distortion to the $Mn_{12}O_{12}$ core is necessary in order to flip JT axes. However, the flipping of a JT axis, by itself, is unlikely to cause a significant enough reduction in the dominant axial anisotropy, $D$, to explain the observed reduction in $U_{eff}$. Clearly, other factors must be considered. For example, a significant structural distortion will likely impact the exchange pathways within the molecule, possibly stabilizing a different spin ground state (or reducing the proximity of the $S = 10$ state to excited spin states). In addition, one may expect a significant projection of the $Mn^{III}$ single-ion anisotropy into the hard plane, resulting in a non-negligible rhombic anisotropy term, $E(\hat{S}_x^2 - \hat{S}_y^2)$, that may enhance QMT below the top of the barrier.

In the past few years, it has been recognized that the FR species in typical $Mn_{12}$-acetate samples can have a profound influence on the quantum dynamics associated with the SR majority species. In particular, it was shown by Wernsdorfer that reversal of FR molecules may trigger magnetization avalanches among the majority species.[15] In turn, these avalanche phenomena have themselves become the focus of much recent attention in $Mn_{12}$-acetate.[16,47-50] More dramatically, it has been shown by Morello et al.,[51,52] that the coupled nuclear spin-lattice dynamics is strongly influenced by the FR species in the low-temperature regime where the SR molecules are completely blocked and, thus, should not mediate any coupling between the nuclear spins and the lattice. Therefore, these findings motivated more detailed spectroscopic investigations, as described in this article.

The paper is organized as follows: in the next section (II), we review previous spectroscopic efforts aimed at characterizing FR $Mn_{12}$; in the following section (III), we describe our experimental methods, including a new angle-swept EPR technique which facilitates simple alignment of low-symmetry crystals; we then describe in detail the obtained EPR data in Section IV, followed by further discussion of the results in section V; finally, we conclude with a summary of the paper in Section VI.

## II. Previous spectroscopic measurements on FR $Mn_{12}$

Hysteresis measurements provide important spectroscopic insights. First of all, as mentioned above, hysteresis steps due to QMT have been observed for $Mn_{12}$-acetate crystals which can be attributed to the FR species.[14-16] The spacings between these steps suggest a slight reduction in the molecular $D$ value of order 10-20%. However, these studies do not enable an accurate determination of the magnetization barrier, $U_{eff}$, for several reasons: the ground state spin value associated with the FR species is not well known; higher order axial anisotropy terms that may contribute to the barrier height are also unknown; and transverse (off-diagonal) interactions that may cause tunneling below the top of the barrier are not well determined. Nevertheless, hysteresis measurements have yielded other important information. For example, it is known that the FR species in the acetate have their magnetic easy axes tilted significantly with respect to the axial SR species. In fact, four separate FR easy axis directions are found, tilted ~10° with respect to the crystal $c$-axis in four orthogonal planes, consistent with the tetragonal space group.[14] Thus, it appears that the FR species occupy sites in the bulk of the crystal which retain the average tetragonal symmetry. However, one should not think of the tilting as resulting from an actual rotation of the molecule. It is simply a reorientation of the magnetoanisotropy tensors caused by the abnormally oriented JT axes,[53] of which there are presumably four equivalent sites in the molecule (related by the tetragonal symmetry) which can flip.

In addition to the acetate, hysteresis studies have been reported for other $Mn_{12}$ crystals for which the majority species are FR.[24,25] One again finds in these cases that the spacings between the strongest steps are quite similar to those observed in the acetate, indicating that the molecular *D* values associated with the ground spin states may be rather similar. However, a more complex behavior is observed at elevated temperatures, suggesting that there could be low-lying excited spin states that influence the QMT in the thermally activated regime.[25]

Detailed electron paramagnetic resonance (EPR) and inelastic neutron scattering (INS) measurements have been complicated by many factors. For the acetate, the challenge involves separating weak signals due to the minority FR species from an already complicated spectrum associated with the SR species.[54-57] This is only really possible at low temperatures where one observes just a few well resolved ground-state transitions from the various different species in the crystal (not only the JT isomers, but also the disordered solvent isomers). Thus, such studies provide little new information compared to hysteresis measurements, i.e. one obtains only the zero-field splitting (ZFS) between the ground and first excited states within the lowest-lying spin multiplet. Consequently, this does not enable a reliable estimate of the magnetization barrier for all the same reasons mentioned above in connection to hysteresis studies. Nevertheless, we note that spectroscopic signatures of FR species in $Mn_{12}$-acetate have been previously reported,[55,56] including one of the earlier single-crystal EPR studies.[54] These reports indicate a ZFS, $\Delta_o \approx 220$ GHz, between the ground and first excited state, i.e. ~25% smaller than the SR species.

More detailed magnetization and EPR studies have been reported for two related complexes: $[Mn_{12}O_{12}(O_2CC_6H_5)_{16}(H_2O)_4]\cdot 2C_6H_5CO_2H$ ($Mn_{12}$-benzoate[13,33]) and $[Mn_{12}O_{12}(O_2CC_6H_4\text{-}2\text{-}CH_3)_{16}(H_2O)_4]\cdot CH_2Cl_2\cdot 2H_2O$ ($Mn_{12}$-methylbenzoate[38]). Like the acetate, the benzoate exhibits mixtures of FR and SR species, with ratios that vary from batch to batch. The structure of this particular benzoate is rather complex, even in its relatively pure FR form: the JT isomerism involves a compression instead of an elongation for one of the eight $Mn^{III}$ atoms;[33] and the structure contains two differently oriented molecules in the unit cell. Nevertheless, important insights have been obtained from combined magnetic and EPR measurements. In particular, it is again found that the magnetic easy axes associated with the FR species are tilted considerably (~12°) with respect to the molecular axis. In addition, evidence for anisotropy within the hard plane is presented. Surprisingly, the magnitude of the *D* value obtained for this benzoate is rather low, i.e. 0.45 K compared to 0.67 K for the acetate. In contrast, EPR studies for the methylbenzoate complex, which has two abnormally oriented JT axes, give a *D* value which is quite similar to the acetate.[38]

Overall, the available studies of FR $Mn_{12}$ paint a complex picture. It is clear that the JT isomerism plays a crucial role, resulting in a significant lowering of the symmetry of the $Mn_{12}O_{12}$ core. This, in turn, likely results in several combined factors that may lead to faster relaxation, i.e. a reduction in *D*, a reduction in *S* (or the proximity of the *S* = 10 state to excited spin multiplets), and faster tunneling due to significant anisotropy within the hard plane. The low symmetry of the FR complexes greatly complicates matters. We therefore set out to perform detailed angle-resolved single-crystal high-frequency EPR (HFEPR) studies of the FR-$Mn_{12}$tBuAc complex. When handled correctly, high-quality crystals are obtainable that do not contain significant quantities of the SR molecules. In addition, the $[Mn_{12}O_{12}(O_2CCH_2Bu^t)_{16}(H_2O)_4]\cdot CH_2Cl_2\cdot MeNO_2$ FR form possesses only one molecular orientation, thus facilitating interpretation of the spectra.

## III. Experimental details

The low symmetry of the FR species is the main factor contributing to the complexity of single-crystal HFEPR studies. First of all, crystals form in a variety of morphologies having irregular shapes. What is more, the molecular orientations are rarely related in a simple way to the crystal morphology. One obvious solution involves X-ray screening of a few samples in order to determine the orientations of various crystal faces and then to make attempts to estimate the orientations of the principal magnetic axes relative to these faces. However, numerous such attempts have been made in our group, which have largely been unsuccessful for a variety of reasons, all of which we understand. First of all, even if one could precisely align a sample, one cannot know *a priori* the orientations of the principal magnetic axes. In particular, the hard/medium plane IS NOT coincident with the plane of the $Mn_{12}$ molecule, neither is the easy axis perpendicular to this plane. Second, all of the FR samples that we have handled contain highly volatile solvents of crystallization. Thus, one does not have the luxury of carefully aligning a crystal prior to its study. Indeed, we have found it essential to immediately cover crystals in silicone grease or paratone oil in order to prevent rapid solvent loss. In some cases, crystals can turn to powder upon exposure to air in a matter of minutes, or even seconds.[58] For this reason, careful *ex-situ* sample alignment is essentially impossible. We instead take advantage of a unique HFEPR spectrometer developed formerly at the University of Florida that enables *in-situ* two-axis rotation of the sample, i.e. we use the HFEPR spectrometer to align the sample. This instrument is described in detail elsewhere.[59] However, we recount several important features that were essential to the present investigation.

Two-axis rotation is achieved using the combination of a cylindrical $TE_{01n}$ cavity ($n = 1,2,...$) with a rotating end-plate, and a 7 tesla split superconducting solenoid with a horizontal field. The split-coil magnet design allows for rotation of the horizontal cylindrical cavity about a vertical axis. Coupling of microwaves to and from the cavity is achieved via waveguides attached rigidly to the cylindrical side walls. Orientation of the sample about a 2nd (horizontal) axis is achieved by controlled rotation of the cavity end-plate. This mechanism does not affect either the tuning of the cavity, or the coupling. This instrument, which was originally designed for a completely different purpose,[60] allows transmission measurements as a function of the field orientation relative to the axis of the cavity, i.e. angle-swept HFEPR. In this mode, the rotating end-plate is clamped while the entire EPR probe is rotated smoothly about the vertical axis using an automated stepper motor mounted at the top of the magnet cryostat; the stepper motor provides very fine angle resolution (<0.05°). Subsequent adjustment of the cavity end plate (whilst not recording data) enables control over the plane of rotation. In this way, one can map out the microwave response of a sample over a full $4\pi$ steradians.

In addition to the above mechanical features, the following details of the HFEPR spectrometer prove to be crucial. Due to the significant ZFS associated with a SMM, high frequencies are essential. The fundamental mode of the rotating cylindrical cavity is $f_{TE011}$ ≈ 51.8 GHz (the exact frequency depends on loading and on the temperature). However, the cavity performs optimally on many higher harmonic modes up to around 110 GHz,[61] and continues to work very well for frequencies up to ~450 GHz, even though the mode structure is not well characterized for frequencies above ~150 GHz. Broad frequency coverage is achieved using a millimeter-wave vector network analyzer (MVNA[61]) with an associated external source (ESA) option (a tunable Gunn diode); this spectrometer enables phase sensitive measurements with exceptionally high stability and signal-to-noise characteristics. The sample/cavity temperature is controlled by means of a cold helium gas flow cryostat belonging to a Quantum

Design Physical Property Measurement System (PPMS). The MultiView software associated with the PPMS enables remote control over the sample temperature, the magnetic field strength and its orientation relative to the horizontal axis of the cavity.

For a variety of reasons, sample orientation is best achieved using frequencies close to the fundamental mode of the cavity (50-90 GHz range). These reasons include: optimum sensitivity and signal-to-noise, the ability to use flexible coaxial transmission lines to couple the EPR probe to the MVNA,[61] and the fact that the $Mn_{12}$ EPR spectrum for this frequency range is incredibly sensitive to the field orientation when it is close to alignment with the hard plane.[43,53] However, a completely blind search for the hard plane using conventional field-swept measurements is potentially very time consuming, involving fully energizing the magnet many times for many different field orientations. Indeed, this may easily take several weeks of round-the-clock operation of the spectrometer and the consumption of several hundred liters of liquid helium. For this reason, we have found that a first iteration using the angle-swept method provides a very good idea as to the location of the hard plane. For highly anisotropic species such as $Mn_{12}$, the low-temperature ($T < 25$ K) spectral weight shifts very rapidly to high fields as the field orientation approaches the hard plane (for a detailed explanation, see ref. [53]). One can, therefore, set the magnet into persistent mode at a high field value and locate the approximate hard plane direction by performing wide angle-sweeps (span ≈ 300 degrees) for different planes of rotation (Figs. 2 and 3). In this way, one observes a series of sharp EPR absorptions each time the field cuts through the hard plane, as seen in Fig. 2 and the insets to Fig. 3b. Importantly, if the field is high enough, these absorptions will only be observed within 10-15 degrees of the hard plane. Each angle-swept measurement takes a matter of minutes. Thus, a complete mapping of the hard plane can be achieved in 1-2 hours (Fig. 3). Subsequent refinements can be made by performing a series of field-swept measurements close to the estimated hard plane (Fig. 3b and Fig. 5, section IV); these may be easily automated and run overnight using the MultiView software.

Finally, as already mentioned, great care was taken to avoid solvent loss from the sample. All of the presented data for FR-$Mn_{12}$tBuAc were obtained for a single plate-like crystal with the two large surfaces having a rhomboidal shape (dimensions ~0.8×0.8×0.3 mm$^3$). The sample was transferred rapidly (in just a few seconds) from the mother liquor into Paratone oil, then mounted with one of its large faces flush against the cavity end plate, roughly mid-way between the axis of the cavity and the side walls. Care was taken to remove excess oil from the end-plate prior to sealing. The cavity was then transferred to the PPMS and cooled under 1 atm. of helium gas to a temperature of ~200 K within less than 10 minutes of removing the sample from its mother liquor. Once below this temperature, solvent loss is no longer a concern, enabling a normal slow cooling cycle under vacuum. Nevertheless, an additional check was performed upon completion of the experiments. The cavity was removed from the PPMS at a temperature of ~230 K. The sample was then extracted and transported rapidly to the University of Florida Chemistry Department where it was loaded into a SQUID magnetometer and again cooled quickly. Ac measurements were then performed to check whether a significant fraction of the sample had converted to the SR species (Fig. 4). Remarkably, no detectable HT $\chi''$ signal was found, indicating that no appreciable conversion had taken place. We note also that the frequency dependence of the LT $\chi''$ peak gives a value for the barrier, $U_{eff} = 43(2)$ K, which is in excellent agreement with previous investigations.[25]

## **IV. Experimental results**

Angle-swept EPR spectra for FR-Mn$_{12}$tBuAc, obtained at $T = 20$ K, $f = 61.980$ GHz (the TE$_{012}$ mode with $Q = 13,000$) and a field of 3.794 tesla, were presented in the previous section (Figs. 2 and 3). The first point to note is that the background signal has the rough form of a figure-of-eight when viewed on a polar plot. As a first approximation, this is exactly what one expects for an axial system as one rotates the field in a plane inclined to the hard plane, i.e. a two-fold symmetry (easy-hard-easy...). The nodes occur when the field is closest to the hard-plane where the absorption is strongest. The reason why the figure-of-eight pattern rotates as the plane of rotation ($\phi$) is varied is due to the fact that the sample's magnetic hard ($xy'$) plane is inclined significantly (by an angle of ~39°—*vide infra*) relative to the cavity end-plate ($xy$ plane). This may be understood with the aid of Fig. 3a where one sees that, as $\phi$ is varied, the angle, $\theta$, at which the field intersects the $xy'$ plane varies also (red dots at the edge of the blue disk).

Upon closer inspection of Figs. 2 and 3, one can tell from the asymmetry in some of the data that FR-Mn$_{12}$tBuAc possesses anisotropy within its hard plane. Most apparent is the distorted/skewed nature of the figure-of-eight background for certain $\phi$ angles. More subtle is the asymmetry in the sharp resonances observed close to the nodal locations associated with the figure-of-eight, as illustrated more clearly in the insets to Fig. 3b. In particular, the data obtained at $\phi = -7.8°$ (Fig. 2a and upper left inset to Fig. 3b) exhibit very pronounced asymmetry, with the sharp resonances occurring mostly to one side of the node (higher $\phi$ side). In contrast, the data obtained at $\phi = 80°$ (Fig. 2f and lower right inset to Fig. 3b) are rather symmetric. These differences can be traced to the fact that, in general, the plane of field rotation is inclined at an acute angle relative to the sample's magnetic hard plane. Thus, not only does the out-of-plane $z´$ component of the field vary upon rotation, but also the in-plane ($xy´$) component. The dependence of the EPR spectrum on the $z´$ component of the field *must* be symmetric about the hard plane. However, this need not be so for the $xy´$ component if there is any magnetic anisotropy within the hard plane, because the field intersects the hard plane at different locations for different planes of rotation [see red dots in Fig. 3(A)]. Thus, in general, there is no reason to expect the symmetries associated with the $z´$ and $xy´$ components of the field to be commensurate. Nevertheless, for azimuthal ($\phi$) angles close to 90°, the field makes an approximately perpendicular cut through the hard plane [see Fig. 3(a)]. In this case (Fig. 3b lower right inset), the variation of the $xy´$ component of the field is minimal, hence the more-or-less symmetric appearance of the data.

Given the apparent hard-plane anisotropy, we set out to characterize this behavior in more detail by recording conventional field-swept spectra for different field orientations within the hard plane. However, previous studies have shown that such spectra for $Mn_{12}$ complexes with similar spin and uniaxial anisotropy ($D$) are extremely sensitive to the field alignment relative to the hard plane, i.e. the spectrum changes completely with as little as 1 degree of misalignment.[53] Unfortunately, the hard plane cannot be determined with this precision on the basis of the angle-swept measurements discussed above, particularly given a small backlash associated with the mechanism driving the end plate rotation.[59] We thus collected field-swept data over narrow angle ranges (in 1 degree steps) either side of the estimated hard-plane orientations deduced on the basis of the angle-swept measurements. While this may seem like a step backwards, we note that the number of such measurements was reduced dramatically compared to a blind search for the hard plane, potentially saving many weeks of spectrometer time. Fig. 5a displays a series of such field sweeps recorded at a temperature of 20 K and a frequency of $f = 61.980$ GHz, spanning a range of roughly 4 degrees either side of the hard-plane direction for a plane of rotation corresponding to $\phi = 62.5$ degrees. Fig. 5b shows plots the positions of the observed peaks as a function of the PPMS angle, $\theta$, over a slightly extended range. The peaks have been labeled according to a scheme developed for Mn$_{12}$Ac,[53] and under the assumption that the spin ground state for the FR-Mn$_{12}$tBuAc is also $S = 10$ (see discussion further below). For this end-plate orientation, the hard plane was located at $\theta = 34\pm1$ degrees.

By repeating measurements such as those displayed in Fig. 5 for different end-plate orientations, the hard-plane can be mapped out with great precision (open squares in Fig. 3b). From a fit to these data, one can conclude that the sample's hard plane is inclined $39\pm1$ degrees relative to the cavity end plate, i.e. neither the hard plane nor the easy axis bear any simple relation to the faces of the crystal. This is not unexpected on the basis of the crystal structure, which reveals that the molecules do not align in any simple way with the principal crystallographic directions.[25] Spectra corresponding to different field orientations within the hard plane are displayed in Fig. 6(a). The data reveal a very strong two-fold (rhombic) anisotropy within the hard plane, as illustrated in Fig. 6(b) where the various peak positions are plotted as a function of the angle $\phi$—the curves are simply guides to the eye, and the peak coloring and labeling is the same as in Fig. 5 and will be used throughout the remainder of the paper. The observed angle-dependence indicates that the rhombicity is commensurate with the measurements displayed in Figs. 2 and 3. We note that there is no particular reason why this has to be the case. Nevertheless, the rhombicity appears to follow the tilting of the hard plane away from the flat surfaces of the crystal. Consequently, the medium axis is parallel to the flat surfaces of the crystal, while the hard axis is maximally inclined at 39 degrees relative to the flat sample surface [see also Fig. 3(a)].

We defer detailed discussion of the hard plane anisotropy to Section V and instead shift focus to measurements with the field aligned in the easy-axis direction, from which it is straightforward to deduce the axial ZFS associated with the FR-Mn$_{12}$tBuAc complex. The easy axis can be found by inspection of Fig. 3: the end-plate orientation is set to $\phi = 90^\circ$, and the PPMS angle rotated 90$^\circ$ past the hard plane orientation, i.e. $\theta = 90^\circ + 39^\circ = 129^\circ$. A series of spectra obtained at different high frequencies is displayed in Fig. 7. Both the dynamic range (signal-to-noise) and sensitivity of the spectrometer are significantly reduced at 300 GHz compared to the lower frequencies used for the hard plane measurements, thus explaining the reduced quality of the data in Fig. 7 when compared to Figs. 5 and 6. Nevertheless, it is still possible to identify several clear resonance branches, as indicated by the dashed lines in the figure. The positions of the strongest peaks were tabulated and are plotted versus frequency in

Fig. 8. The red and blue lines represent two separate simulations which superimpose upon the data quite well. The reason for the two simulations becomes apparent upon examination of the temperature dependence of the high-frequency easy-axis data, as we now explain.

A representative set of temperature dependence data are plotted in Fig. 9 for a frequency of 288 GHz, which is slightly above the largest ZFS and also gives the highest quality spectra for this frequency range. Below 10 K, all of the EPR intensity collapses into the lowest field resonance, corresponding to the transition with the largest ZFS of ~275 GHz. This clearly corresponds to the transition from the ground state of the system, i.e. $m_S = -10$ to $-9$, within the presumed $S = 10$ ground state. The first point to note is that the ZFS is less than for high-symmetry SR Mn$_{12}$, for which the known examples give values in the 300-307 GHz range.[34,35,62] Consequently, one may conclude that the reduction in ZFS is, at most, 10% for FR-Mn$_{12}$tBuAc relative to SR Mn$_{12}$, in spite of the fact that the barrier is ~40% lower.

Upon raising the temperature above 10 K, one can discern a series of weak transitions in the 1-3 T range, followed by a stronger resonance at ~4 T. On the high-field side of the 4 T transition, one sees additional weak resonances. This pattern of intensity is not typical for a good SMM, where the peak intensities would typically decrease monotonically from low to high field. We thus believe that the 4 T transitions belongs to a low-lying excited spin multiplet that competes for thermally activated population/intensity with the ground state multiplet. Further clues may be obtained be reexamining the frequency-dependent plots in Fig. 7. For frequencies well above 288 GHz, the resonance branch corresponding to the 4 T peak in Fig. 9 (red data points in Fig. 8) vanishes. However, one begins to see a 2nd strong peak to the right of the ground state resonance at the highest frequencies. The position of this peak corresponds well to what one would expect for a transition from the first excited state within the $S = 10$ ground state multiplet, i.e. $m_S = -9$ to $-8$; this and the $m_S = -10$ to $-9$ transition are represented by the blue data points in Fig. 8. Subsequent transitions ($m_S = -8$ to $-7$, $-7$ to $-6$, etc..) are not clearly observed (see below). Nevertheless, the observation of two transitions enables estimation of the 2$^{nd}$ and 4$^{th}$-order axial ZFS parameters: assuming $S = 10$ (*vide infra*), $D = -0.42(1)$ cm$^{-1}$ and $B_4^0 = -2.2(3) \times 10^{-5}$ cm$^{-1}$. These values are only marginally different from those obtained for high-symmetry Mn$_{12}$ complexes,[34,35] with the slightly lower $D$ value mainly accounting for the reduced ZFS (~275 GHz as opposed to ~300 GHz).

All of the above trends may be reconciled if one considers a low lying excited state with a spin value that is less than that of the ground state. The fact that only the $S = 10$, $m_S = -10$ to $-9$ transition is clearly seen at the lower frequencies suggests that the excited state almost completely overlaps the $S = 10$ ground state, i.e. the separation between the lowest levels in the two spin multiplets is of the order (or even less than) the ZFS within the ground state. If one then assumes that the spin value of the excited state is appreciably lower than that of the ground state, say S = 7, the separation between the lowest levels in the two multiplets will increase rapidly with field so that the intensity associated with the excited state will disappear at higher fields/frequencies, giving way to more and more transitions associated with the ground state. This is exactly what is found experimentally. In order to illustrate this idea, we have superimposed the Zeeman energy-level diagrams for a spin $S = 10$ state and a $S = 7$ state in Fig. 10. It is important to emphasize that the inclusion of the $S = 7$ state is purely schematic, and should not be taken too literally. Only 2nd and 4th order ZFS axial terms were included in these simulations so as to obtain agreement with the data in Fig. 8 (red and blue lines), i.e. no transverse anisotropy was considered at this stage, which is an over simplification. The ZFS parameters used for the $S = 7$ state were: $D = -0.46$ cm$^{-1}$, $B_4^0 = +8.3 \times 10^{-6}$ cm$^{-1}$ and $g_z = 2.00$. These were chosen so as to give good agreement with the excited state transitions (red data points) in Fig. 8; we note that the larger $D$ value, as compared to the $S = 10$ state, is not unexpected, because $D$ tends to increase as $S$ decreases.[63] Finally, the zero-field separation between the lowest levels associated with the two spin multiplets was chosen to be ~7 K.

It should be emphasized that the quoted spin values are not that well determined due to the limited number of observed transitions. This is true even for the $S = 10$ state, because only two transitions are observed; though we note that a value of $S = 10 \pm 1$ has been deduced independently from magnetic measurements.[25,64] One typically needs to see on the order of $S$ resonances in a spectrum in order to obtain a truly accurate value for the total spin. It is also likely that many other spin states exist just above the first excited spin multiplet, resulting in a huge density of spin states at energies just 10-20 K above the ground state. This density of states will compete with the lowest lying levels for thermally activated spin population, and likely explains why just a few broad resonance peaks are observed. The broad nature of the peaks may also be a sign of increased spin relaxation between the many low-lying states. In contrast, the hard plane spectrum is well resolved at high fields and low frequencies. The high field provides an advantage in terms of better separating the ground spin multiplet from the higher-lying states, and the lower frequencies provide enhanced signal-to-noise.[61] For this reason, the hard plane spectra are cleaner and easier to interpret. Nevertheless, the easy axis data provide a direct measure of the ZFS, from which one can deduce the axial spin Hamiltonian parameters $D$ and $B_4^0$, thus reducing the number of parameters needed to fit the hard plane data. These easy axis measurements also provide a further confirmation of the alignment of the sample.

Although purely schematic, Fig. 10 provides many useful insights concerning the properties of FR-Mn$_{12}$tBuAc. In fact, one can even begin to gain a qualitative understanding of low temperature hysteresis measurements,[25] which exhibit many additional thermally activated QMT steps at elevated temperatures (1.4-4.4 K), in-between the principal evenly spaced QTM steps observed in the pure quantum regime. As originally pointed out,[25] the additional steps likely involve tunneling between states associated with different spin multiplets. First of all, the ZFS parameters obtained from this study reproduce the positions of the pure QMT steps very well, i.e. the lowest crossings between the blue energy levels in Fig. 10, which are separated by ~0.5 T. Not far above these level crossings, one finds many additional crossings between the ground and excited spin multiplets (red and blue levels). These can account well for the thermally activated QTM steps observed in the hysteresis experiments.[25] It should be emphasized that no attempt has been made here to obtain quantitative agreement with the hysteresis studies. We note that recent studies of a Mn$_6$ SMM have enabled such quantitative comparisons,[65,66] in part because of the reduced dimension of the Hamiltonian matrix for Mn$_6$ in comparison to Mn$_{12}$.

The studies presented in this section clearly highlight several important basic facts concerning FR Mn$_{12}$: (a) its axial anisotropy is only marginally less than the SR high-symmetry Mn$_{12}$ species; (b) it possesses a considerable magnetic anisotropy within its hard plane; and (c) there exists a very low-lying excited spin state, with a significantly reduced spin value.

## <u>V. Discussion</u>

The primary focus of this study involves establishing the physics behind the effective barrier reduction associated with FR Mn$_{12}$. The fact that the ZFS is only marginally lower for FR Mn$_{12}$, as compared to SR Mn$_{12}$ (~275 GHz versus 300 GHz[34]), suggest that this is not due simply to a reduction in the axial anisotropy ($D$). Even accounting for a possible reduction in the total spin to $S = 9$, the EPR barrier reduction would be only on the order of 25% relative to high-symmetry

Mn$_{12}$ species, in contrast to a reduction in $U_{eff}$ of ~40% found experimentally (Fig. 4). However, it should be emphasized that combined EPR and magnetic[25] studies overwhelmingly favor a spin $S = 10$ ground state, as outlined further below. Consequently, relaxation in the FR species must occur below the top of the theoretical barrier deduced on the basis of EPR measurement of the axial ZFS parameters. Hereon, we refer to this theoretical maximum barrier as the EPR barrier, $U_{EPR}$, in contrast to the kinetic or effective barrier, $U_{eff}$, deduced from ac susceptibility data such as those displayed in Fig. 4.[34,67]

Clearly, the presence of strong transverse ZFS interactions can account for the effective barrier reduction. These interactions will lead to a strong mixing of spin projection states, thereby providing highly efficient relaxation pathways at energies considerably below the top of the EPR barrier, as illustrated schematically in Fig. 11. We note that previous studies have clearly demonstrated that disorder induces weak random transverse anisotropic interactions locally,[40-43,68] and that this results in a small, yet systematic effective barrier reduction.[34] These situations are illustrated in Fig. 11, i.e. weak disorder causes tunneling slightly below the top of the EPR barrier, while samples with intrinsically low symmetry (Jahn-Teller isomers) tunnel well below the top of the barrier.

The data displayed in Fig. 6 reveal a dramatic two-fold dependence on the orientation of the field within the hard-plane, suggesting a very significant rhombic anisotropy. Fig. 12 displays attempts to simulate the combined hard-plane data on the basis of the frequency- and field-orientation-dependence of the α resonances. These simulations were additionally constrained by the easy-axis data in Fig. 8. Reasonable agreement is obtained for all frequencies and field orientations using the following parameters (in addition to the axial ones quoted above): $S = 10$, $E = -0.072$ cm$^{-1}$ (the sign is undetermined, and chosen here so that medium direction coincides with $\phi = 0$), $B_4^2 = +1.5 \times 10^{-5}$ cm$^{-1}$ and $B_4^4 = -3 \times 10^{-5}$ cm$^{-1}$. The solid curves in Fig. 12 are simulations generated using these parameters together with the previously obtained axial ones ($D$ and $B_4^0$). The dominant parameter after $D$ is the rhombic term $E$ ($\approx D/6$); this degree of rhombicity is comparable to that of the classic biaxial SMM Fe$_8$.[69] The two fourth-order terms primarily serve to provide incremental improvements in the simulations, i.e. setting both to zero only marginally affects the simulations. Nevertheless, we note that comparable $B_4^4$ values are found for the SR Mn$_{12}$ complexes.[37,40-43,45,46] In addition, given the low symmetry of SR Mn$_{12}$ and the presence of a very low-lying excited spin state, it is quite reasonable to expect a comparable rhombic 4$^{th}$ order term.[70]

Efforts to obtain improved simulations do not serve any useful purpose, because the applicability of the giant spin description is questionable for a molecule that clearly possesses such a low-lying excited spin multiplet (see Fig. 10). Recent work has demonstrated that multiple higher order terms (> 4$^{th}$ order) in the giant spin expansion become significant in these situations.[37, 70] Consequently, one cannot really be certain as to which is the correct parameterization. Nevertheless, the simulations in Fig. 12 indicate that the hard-plane angle-dependence is primarily influenced by the value of $E$; the $B_4^4$ and $B_4^2$ parameters were added simply to improve the overall agreement with the data. The take-home message is that the values of $D$ and $E$ should be considered quite reliable (likely to within ~2-3% for $D$ and ~10% for $E$, assuming $S = 10$), and that it is these two parameters that primarily govern the low-temperature relaxation dynamics: $D$ sets the scale for the classical barrier, $U_{EPR}$; $E$ then controls the effective barrier reduction due to tunneling. We note here that attempts to simulate *all* available EPR and magnetization data on the basis of a spin value other than $S = 10$ yielded far less satisfactory results, lending considerable weight to the assertion that the ground state does indeed correspond to $S = 10$.

We conclude this section by demonstrating that the above parameterization can account quantitatively for the effective barrier reduction in FR-Mn$_{12}$tBuAc. Fig. 13 illustrates the mixing of $m_S$ levels that occurs as a result of the significant transverse anisotropy (primarily $E$), inferred on the basis of the HFEPR measurements. In the absence of any transverse terms, the $m_S$ [projection of $S$ along the molecular easy- ($z$) axis] quantum number is exact. In this situation, each eigenvector is characterized by a single $m_S$ component (with coefficient $C_m = 1.00$), and the eigenvalues lie on the blue curve in Fig. 13, at the locations of the blue dots; this curve was generated using the published 2$^{nd}$ and 4$^{th}$ order axial parameters for SR-Mn$_{12}$tBuAc. The results for FR-Mn$_{12}$tBuAc are represented by the colored islands; the eigenvalues have been referenced to the lowest-lying eigenstates ($m_S = \pm 10$ for the pure case). The first point to note is the reduction in ZFS for the FR complex, i.e. the lowest-lying eigenvalues associated with the FR species (red islands) are reduced in comparison to those of the SR species (blue dots). To first order, this is caused by the ~10% reduction in axial anisotropy. However, the more dramatic effect on the spectrum is caused by the significant rhombic anisotropy, characterized by the large $E$ value (= –0.072 cm$^{-1}$ ≈ $D$/6). Because the rhombic $\hat{O}_2^2$ operator does not commute with $\hat{S}_z$, there is a mixing of states that differ in $m_S$ by ±2. This mixing is very weak at the lowest energies, and strongest at the top of the barrier. For example, one finds only a very small admixture of $m_S = \pm 8$ ($C^2 = 0.003$) into the $m_S = \pm 10$ ground states, while levels that reside at energies of 40K and above contain very significant mixtures of $m_S$ states.

All but one of the energy levels in Fig. 13 occur in pairs (doublets) having a small tunnel splitting which is not discernible in the figure. The exception is the level located at ~49 K, which is a singlet. In this sense, it plays the same role as the unperturbed $m_S = 0$ state. The associated eigenvector is an almost uniform mixture of even $m_S$ states running from –6 to +6, i.e. there is an equal probability of finding a molecule excited to this level on either side of the barrier. In other words, there is no barrier at ~49 K and above. In fact, all eigenvectors above the singlet consist of equal mixtures of positive and negative $m_S$ states; this can easily be seen from the figure, because each of these doublets pair an odd state with an even one. This is not the case for eigenvectors that reside below the singlet level, which consist of either even or odd $m_S$ mixtures; these states also have a definite weighting (asymmetry) towards one or other side of the barrier. Nevertheless, even at ~46 K, there is a very significant $m_S$ 'leakage' across the barrier; the tunnel splitting associated with the 46 K levels is ~500 MHz, and ~40 MHz for the 40 K levels. Therefore, the effective barrier of 43(2) K deduced from ac susceptibility measurements (Fig. 4) is in reasonable agreement with the energy diagram (Fig. 13) derived on the basis of the EPR studies. We note that the inclusion of the higher order terms ($B_4^4$ and $B_4^2$) which were needed to improve the agreement between the simulations and the EPR data in Fig. 12 do not affect the mixing observed in Fig. 13 discernibly, i.e. it is the 2$^{nd}$ order rhombic anisotropy that dominates the effective barrier reduction. Similar trends are observed for the Fe$_8$ SMM, for which the ratio $E/D$ is quite comparable.[71]

Detailed structural, magnetic and HFEPR studies of a wide range of Mn$_{12}$ complexes suggest that they can be grouped into three basic types: (1) Highest symmetry – these are characterized by tetragonal space groups and highly symmetric Mn$_{12}$O$_{12}$ cores (e.g. Mn$_{12}$-bromoacetate[26]); (2) High symmetry – these also possess relatively symmetric Mn$_{12}$O$_{12}$ cores, with no abnormally oriented JT axes, but they crystallize in lower symmetry space groups in which the ligands are not necessarily symmetrically distributed about the magnetic core; and (3) Low symmetry – these crystallize in low-symmetry structures and possess highly distorted $Mn_{12}O_{12}$ cores having one or more abnormally oriented JT axes[18,19,21,23,24] (or even a compression at one of the sites[33]). Type (1) $Mn_{12}$'s exhibit the slowest low temperature relaxation and the highest effective barriers, $U_{eff}$ (~70 K), which are typically close to the classical anisotropy barrier ($U_{EPR}$) one would expect on the basis of the axial ZFS parameters obtained from EPR.[34] Type (2) $Mn_{12}$'s also exhibit slow relaxation ($T_B$ ~ 4 K), albeit with a slight effective barrier reduction (~55–70 K).[28] Finally, type (3) SMMs such as the FR-$Mn_{12}$tBuAc studied here may be characterized as fast relaxing, i.e. they have effective barriers and blocking temperatures that are significantly lower than types (1) and (2). Crystals of the most extensively studied $Mn_{12}$-acetate SMM actually consist of mixtures of all three of these types.[39-42]

We see from the above that the key factor determining the relaxation is the topology of the $Mn_{12}O_{12}$ core. Symmetry lowering at the surface of the molecule leads only to a minor degradation of the SMM properties. The primary source of the magnetoanisotropy in $Mn_{12}$ SMMs comes from the near parallel JT distortions at the eight Mn(III) sites, which gives rise to significant axial single-ion ZFS ($D_{Mn(III)} \approx -3.6$ cm$^{-1}$).[62] For the strictly tetragonal cases, 2$^{nd}$ order transverse anisotropy (finite $E$) is strictly forbidden, i.e. the projection of the 2$^{nd}$ order JT anisotropy onto the hard plane perfectly cancels (though it reemerges in higher orders[70,72]). The cancelation is less effective in situations where there is a minor perturbation of the $Mn_{12}O_{12}$ core due, e.g., to small reorientations of the JT axes caused by an asymmetric distribution of the ligands. In such situations, weak 2$^{nd}$ order transverse anisotropy emerges, leading to weak MQT interactions (see Fig. 11).[34,53,68] For more extreme asymmetric cases, involving a significant reorientation of one or more JT axes, the cancelation is completely ineffective, leading to a projection of most of the anisotropy associated with one or more Mn(III) ions onto the hard plane. This is the situation found in FR type (3) $Mn_{12}$'s.

If one takes the view that the molecular anisotropy is simply the vector sum of the single-ion anisotropies,[71,73] then one can crudely rationalize the ~10% reduction in the molecular $D$ value, and the emergence of a rhombic ZFS parameter, $E \approx D/6$, found for FR-$Mn_{12}$tBuAc, i.e. if one of the eight JT axes flips by 90 degrees, then $D$ should decrease by approximately 1/8, and an anisotropy should appear within the hard plane of order $D/7$. A final consideration centers on the bimodal distribution of effective barriers observed from studies of many neutral $Mn_{12}$ complexes.[28] There are likely two contributing factors. Firstly, for the octahedral geometry, one can only rotate a given O⋯Mn⋯O axis so far. Consequently, as long as the JT distortion involves the axial ligands (those above and below the plane of the $Mn_{12}O_{12}$ core), the projection of the JT anisotropy into the hard plane is likely to be relatively weak. In the case of the JT isomers, the distortion involves the equatorial ligands (those surrounding the periphery of the $Mn_{12}O_{12}$ disk). In this sense, there is no continuum between the two cases, hence the abrupt increase in transverse anisotropy (tunneling) between the type (2) and (3) $Mn_{12}$ SMMs. A second factor concerns the abrupt crossovers that are known to occur in situations involving thermally assisted MQT.[74] This is due in part to the power-law dependence of the various tunnel splittings on the transverse ZFS parameters, in this case $E$. Thus, one could envisage a fairly abrupt transition between the case where tunneling occurs just below the top of the classical barrier, to one in which the relaxation occurs via levels much further below.

## **VI. Summary and conclusions**

We have presented a detailed multi-high-frequency angle- and field-swept electron paramagnetic resonance study of a single crystal sample of $[Mn_{12}O_{12}(O_2CCH_2Bu^t)_{16}(H_2O)_4]\cdot CH_2Cl_2\cdot MeNO_2$ in order to gain a general understanding of the effective barrier reduction and resultant fast relaxation observed in so-called Jahn-Teller variants of $Mn_{12}$. A novel angle-swept HFEPR technique proved essential for accurate alignment of the sample within the HFEPR resonator. This, in turn, enabled a reliable characterization of both the axial and transverse ZFS interactions.

Comparisons between the obtained HFEPR data and published magnetization studies suggest that the ground state spin value remains $S = 10$ for FR-$Mn_{12}$tBuAc, although the presence of very low-lying excited spin states (much lower than for SR $Mn_{12}$) is inferred. The key finding concerns the 2$^{nd}$ order axial and rhombic ZFS parameters $D$ and $E$. A relatively small reduction in $D$ of only ~10% is insufficient to account for the nearly 40% reduction in the effective barrier ($U_{eff}$) to magnetization relaxation deduced from ac relaxation studies. Meanwhile, the presence of a significant rhombic parameter, $E$ ($\approx D/6$), accounts for the enhanced low-temperature relaxation by virtue of under-barrier quantum tunneling resulting from a very significant mixing of spin projection states at energies well below the top of the classical anisotropy barrier. The origin of the strong rhombicity can be traced to an abnormally oriented Jahn-Teller axis associated with one of the Mn(III) ions. We believe that these findings apply quite generally to all FR species of $Mn_{12}$. Indeed, we infer that the minority species found in Mn$_{12}$Ac (and other SR Mn$_{12}$ crystals) also have abnormally oriented Jahn-Teller axes, and that it is the resultant rhombicity that explains the observed fast low-temperature relaxation.

## VII. Acknowledgements

This work was supported by the US National Science Foundation (Grant nos. DMR DMR0804408, DMR0506946 and CHM0910472). We are grateful to Junjie Liu for assistance in generating Fig. 13.

**Figure Captions**

FIG. 1: (Color online) Structures of FR (top) and SR (bottom) $Mn_{12}tBuAc$, illustrating the tilting of one of the eight Jahn-Teller axes associated with the latter into the plane of the molecule. The relevant O13–Mn6–O26 Jahn-Teller axis is labeled in the top figure. Color coding: Mn(III) – blue; Mn(IV) – green; oxygen – red; carbon – gray; $H_2O$ – yellow.

FIG. 2: Angle-swept HFEPR spectra for FR-$Mn_{12}tBuAc$, obtained at $T = 20$ K, $f = 61.980$ GHz (the $TE_{012}$ cavity mode) and a field of 3.794 tesla. The cavity transmission is recorded (radial coordinate) while the PPMS angle, $\theta$ (angular coordinate), is swept continuously by means of a stepper motor associated with the PPMS. The different figures correspond to different planes of rotation, $\phi$, set by the orientation of the cavity end-plate (see also Fig. 3): (a) $\phi = -8°$; (b) $\phi = +9.8°$; (c) $\phi = +27.3°$; (d) $\phi = +44.9°$; (e) $\phi = +62.5°$; (f) $\phi = +80.0°$.

FIG. 3: (Color online) (a) Schematic representation of the experimental (unprimed) and sample (primed) coordinate systems: the *xy*-plane (peach) coincides with the rotatable cavity end-plate, which controls the plane of rotation, $\phi$, of the field; the *xy´* plane (blue) represents the hard plane of the sample, and *z´* its easy axis. In a typical experiment, the field orientation ($\theta$) is scanned at different end plate orientations ($\phi$); the red dots denote intersections between the sample's hard plane and the field rotation plane. (b) Determination of the location of the hard plane as a function of the experimental coordinates $\theta$ and $\phi$. The red open circles represent a first attempt to locate the hard plane on the basis of the angle-swept measurements in Fig. 2; the insets show angle-swept spectra (cavity transmission) as a function of the field orientation, $\theta$, plotted in Cartesian form for two different end-plate orientations ($\phi = -8°$ and $+80.0°$). The black squares were obtained from refined field-swept measurements (see Fig. 5). The blue curve represents the best fit determination of the location of the hard plane.

FIG. 4. (Color online) Main panel: Temperature dependence of the out-of-phase component of the frequency-dependent ac susceptibility, $\chi''$; the frequencies are given in the figure. Inset:

Arrhenius plot of log(frequency) versus the inverse of the temperature, $T_{max}$, corresponding to the maxima in $\chi''$; a linear fit to the data gives the effective barrier, $U_{eff}$, to magnetization reversal.

FIG. 5. (Color online) (a) HFEPR spectra recorded at a frequency of 61.98 GHz and at $T = 20$ K, for different field orientations, $\theta$, either side of the hard plane; the plane of rotation was $\phi = +62.5°$ and the angle step was 1°. (b) Plot of the resonance field positions from (a) versus the angle $\theta$. The peaks are labeled (and color/shape coded) according to a scheme described in Ref. [53]. The estimated location of the hard plane is marked by the dashed line.

FIG. 6. (Color online) (a) HFEPR spectra recorded at a frequency of 61.98 GHz and at $T = 20$ K, for different field orientations within the hard plane (corresponding to different rotation planes, $\phi$, as indicated in the figure); several of the resonances have been identified by the open colored symbols based on the same labeling scheme used in Fig. 5 and Ref. [53]. (b) Plot of the resonance field positions from (a) versus the angle $\phi$; the curves are guides to the eye. The dashed horizontal line corresponds to the hard plane data in Fig. 5.

FIG. 7. (Color online) Easy-axis spectra recorded at several frequencies in the range from 222 GHz (top) to 364 GHz (bottom). The temperature is 20 K, and the dashed lines are guides to the eye.

FIG. 8. (Color online) 2D plot of frequency versus the resonance positions extracted from Fig. 7. The data points have been color/shape coded according to whether the excitations occur within the ground (presumed $S = 10$) state, or an excited state (possibly $S = 7$). The solid and dashed lines superimposed upon the data were simulated using the parameters given in the main text.

FIG. 9. (Color online) 288 GHz easy-axis spectra recorded at different temperatures in the range from 6 K (bottom) to 24 K (top). The dashed lines emphasize the fundamental excitations associated with the ground (presumed $S = 10$) state and the excited state (possibly $S = 7$).

FIG. 10. (Color online) Superimposed Zeeman diagrams corresponding to a spin $S = 10$ ground state and a very low-lying $S = 7$ state. This figure is purely schematic, and can account

qualitatively for the observed temperature dependence (Fig. 9) and quantitatively for the EPR peak positions observed in Fig. 8.

FIG. 11. (Color online) Schematic illustrating the quantum tunneling processes responsible for the effective barrier reduction in $Mn_{12}$ SMMs. For the perfectly axial case, the relaxation is purely classical, and the EPR and effective barriers should be the same ($U_{EPR} = U_{eff}$). Weak transverse ZFS terms, either intrinsic or due to disorder, lead to tunneling between states slightly below the top of the classical barrier [34], i.e. to a small reduction in $U_{eff}$. A significant lowering of the symmetry of the $Mn_{12}$ core due, e.g. to Jahn-Teller isomerism, will result in tunneling significantly below the top of the classical barrier—see main text for further explanation.

FIG. 12. (Color online) (a) 2D plot of frequency (62 – 107 GHz) versus resonance position extracted from a series of spectra recorded at 20 K with the field applied exactly along the sample's medium magnetic axis ($\phi = \theta = 0°$). (b) Plot of 61.98 GHz hard-plane resonance positions versus the angle $\phi$. In both figures, the color/shape coding is the same as used in Figs. 5 and 6, and the solid curves superimposed upon the data represent the best attempt at simulating the combined data sets [Fig. 8 and Figs. 12(a) and (b)]; the parameters used for the simulations are given in the main text.

FIG. 13. (Color online) Schematic illustrating the mixing of unperturbed $m_S$ states caused primarily by the 2$^{nd}$ order rhombic anisotropy. The blue dots and line represent the eigenvalues obtained via a diagonal Hamiltonian (only 2$^{nd}$ and 4$^{th}$ order axial terms) for SR $Mn_{12}$tBuAc, resulting in an energy barrier of ~70 K.[34] The color coded islands represent the mixtures of unperturbed $m_S$ states obtained using the parameters obtained from this EPR investigation (see main text for a more detailed explanation). The islands are colored according to the squared coefficients (probabilities $C_m^2$ associated with each $m_S$ value) corresponding to the normalized eigenvectors; the scale is logarithmic in order to emphasize the mixing. It is important to recognize that all but one of the energy levels consist of a pair of eigenvectors that are not resolved in the figure; the lone singlet level at ~49 K is indicated. Because of this, the diagram appears to be symmetric about $m_S = 0$. However, the levels below 49 K, in fact, consist of one eigenvector with most of its associated probability on the left-hand side of the diagram, and a mirror image on the right-hand side. In contrast, all of the eigenvectors above 49 K are truly symmetric about $m_S = 0$, with each doublet consisting of one eigenvector that is a mixture of even $m_S$ states and another which is a mixture of odd $m_S$ states. This is the reason why the levels above 49 K appear to consist of even and odd mixtures, which would violate the $\Delta m_S \pm 2$ and $\pm 4$, selection rules associated with the 2$^{nd}$ and 4$^{th}$ order transverse operators. However, this is just an illusion created by the fact that each of these levels consists of a superposition of separate even and odd $m_S$ mixtures.

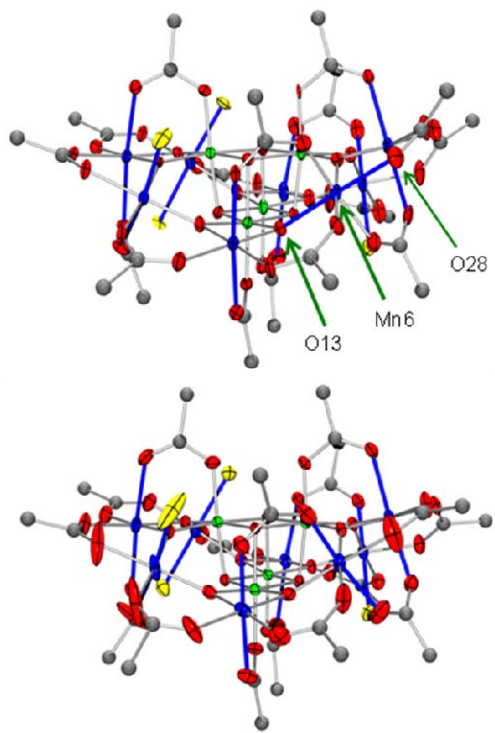

S. Hill et al., Figure 1

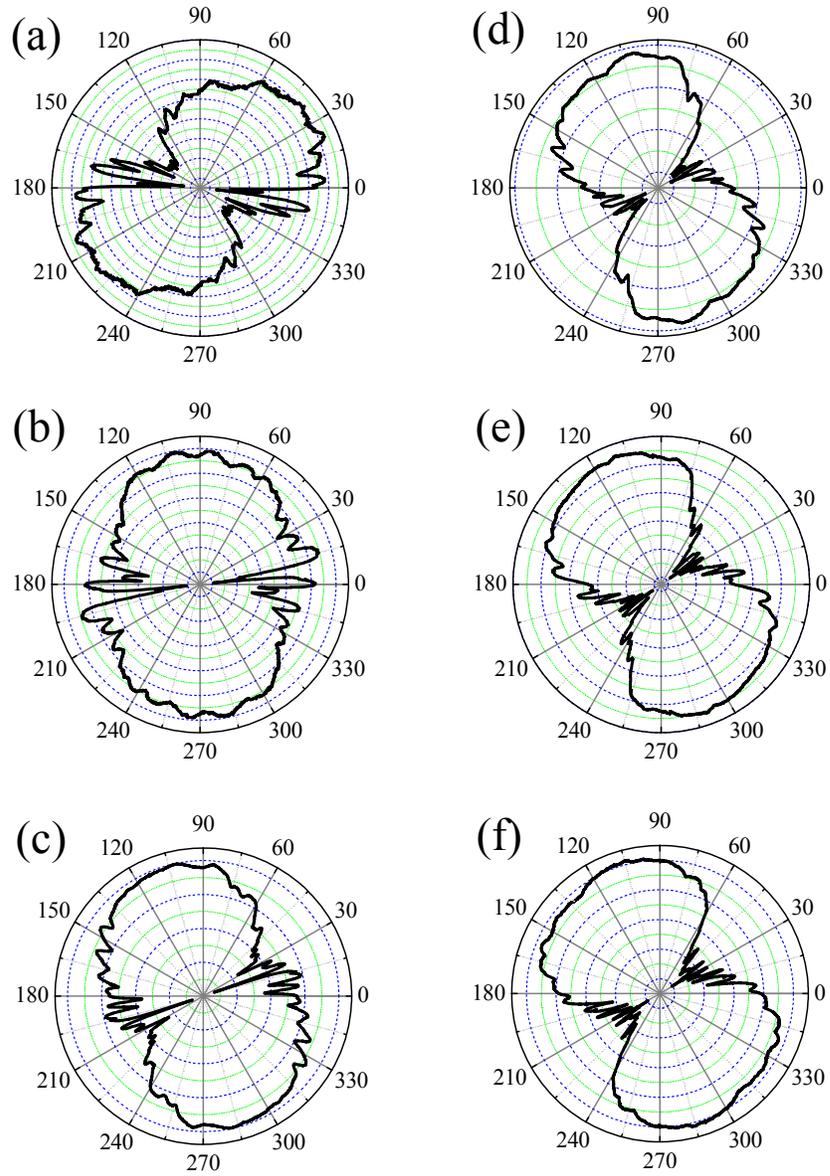

S. Hill et al., Figure 2

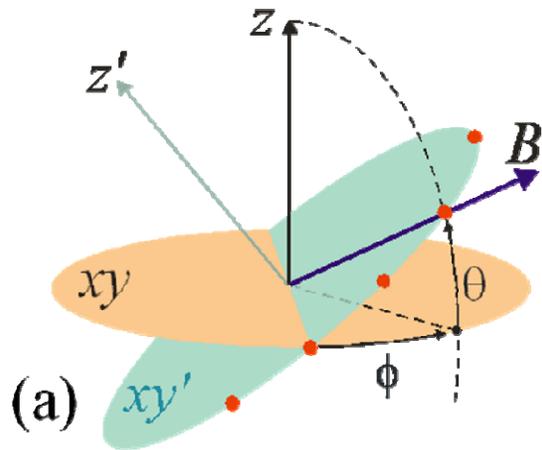

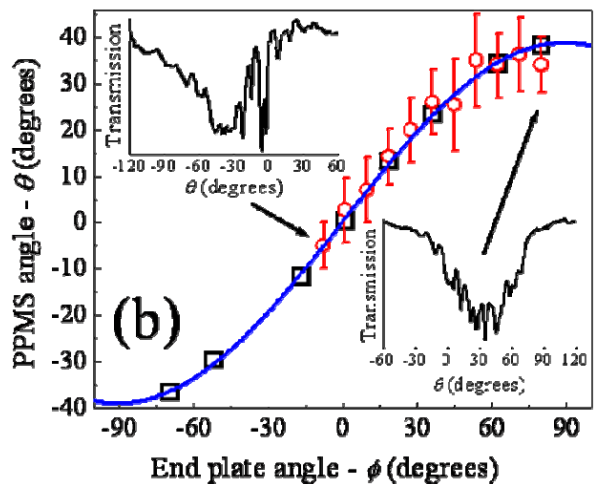

S. Hill et al., Figure 3

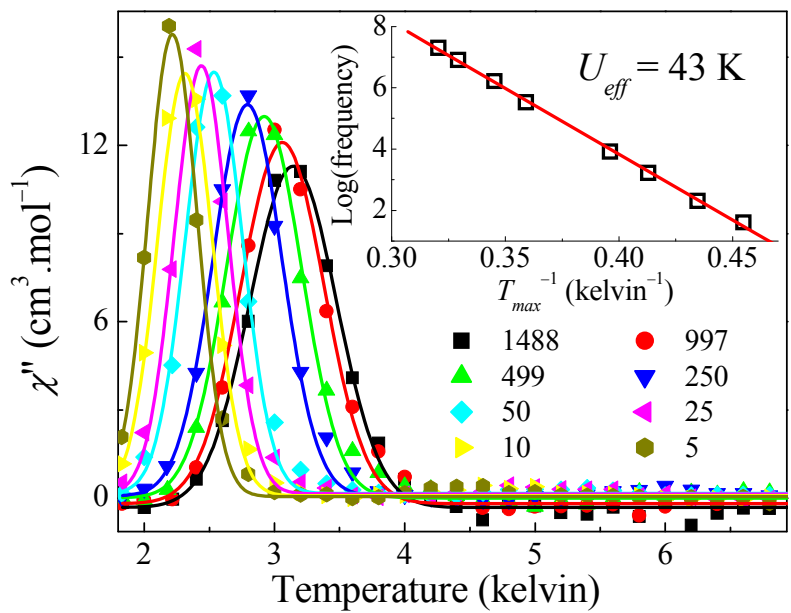

S. Hill et al., Figure 4

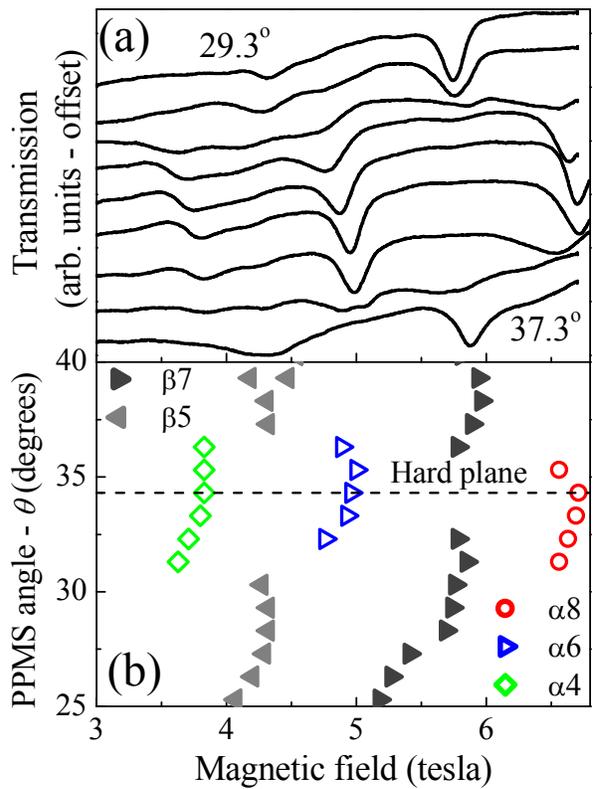

S. Hill et al., Figure 5

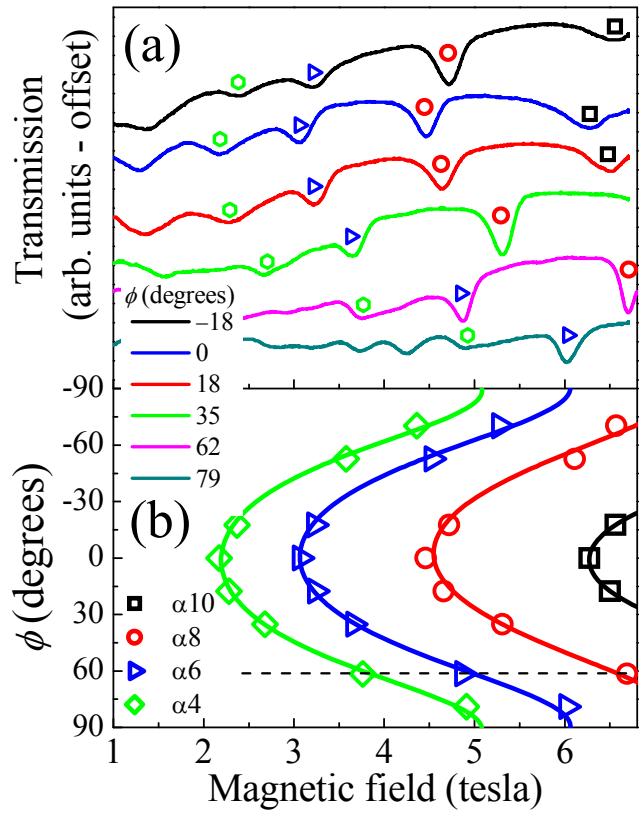

S. Hill et al., Figure 6

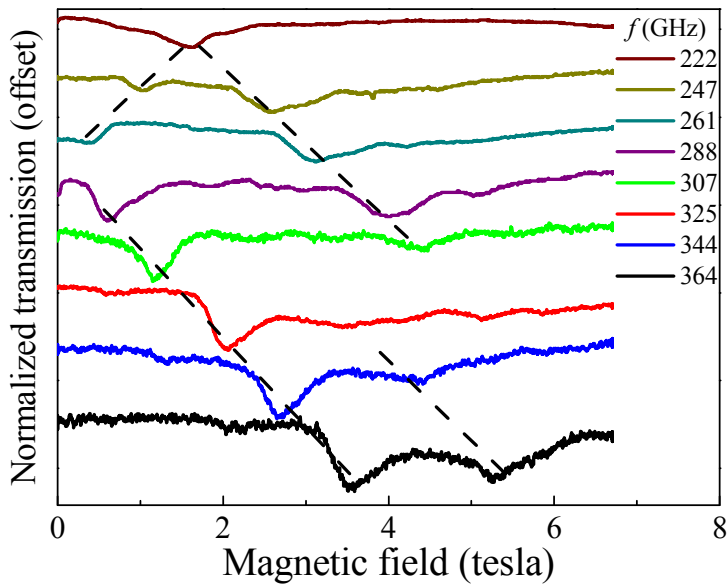

S. Hill et al., Figure 7

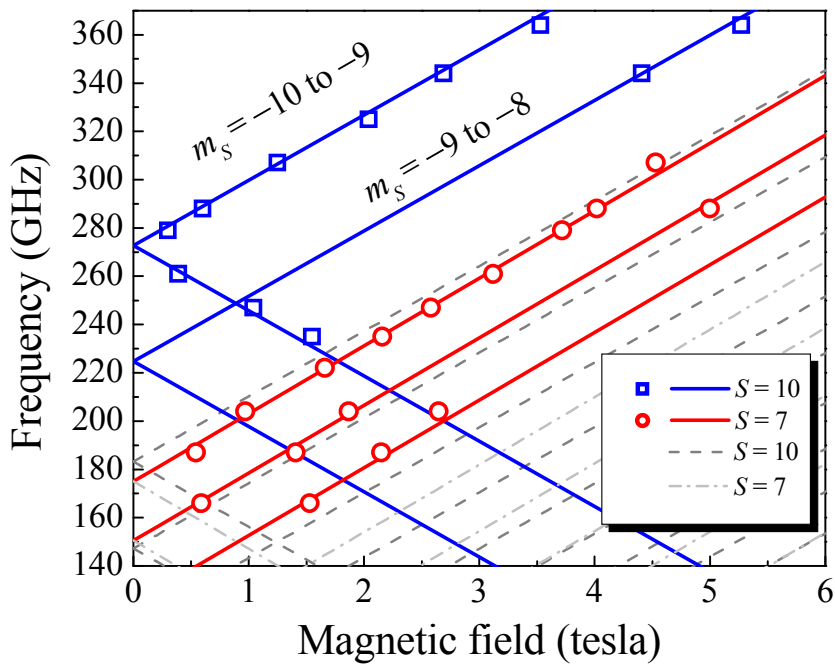

S. Hill et al., Figure 8

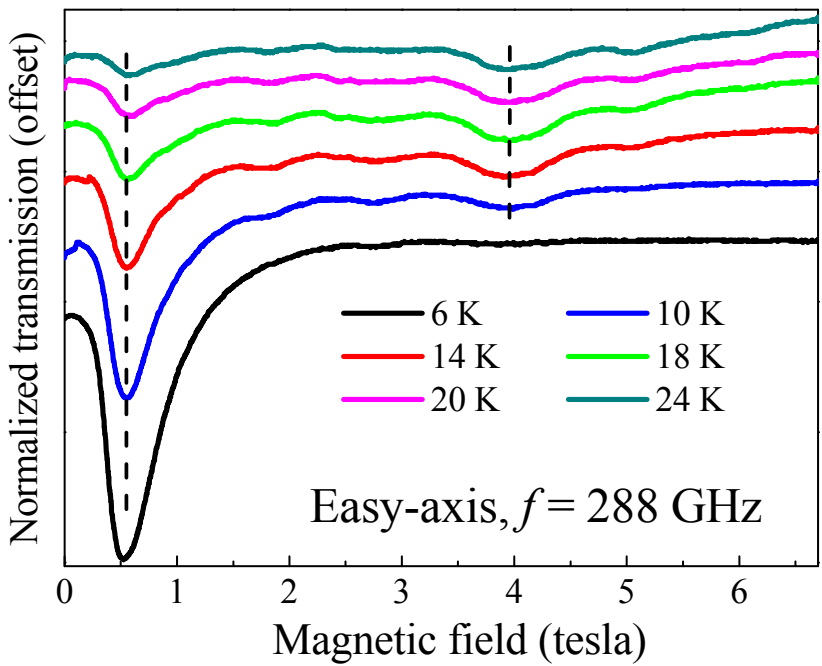

S. Hill et al., Figure 9

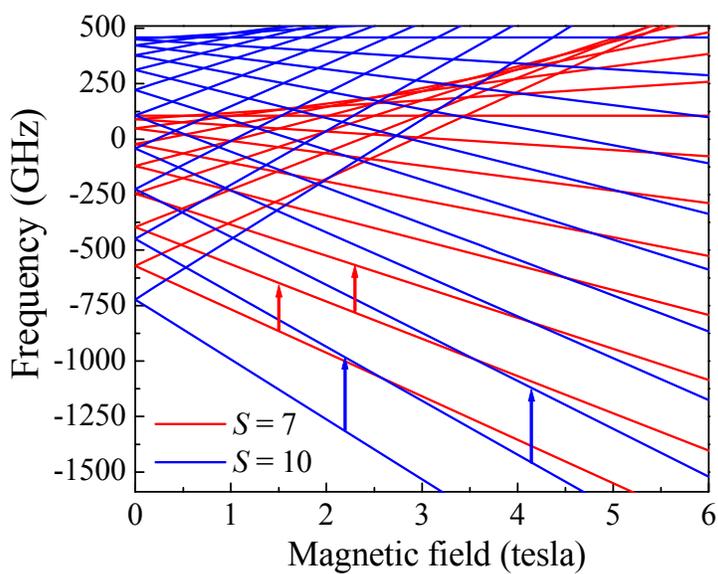

S. Hill et al., Figure 10

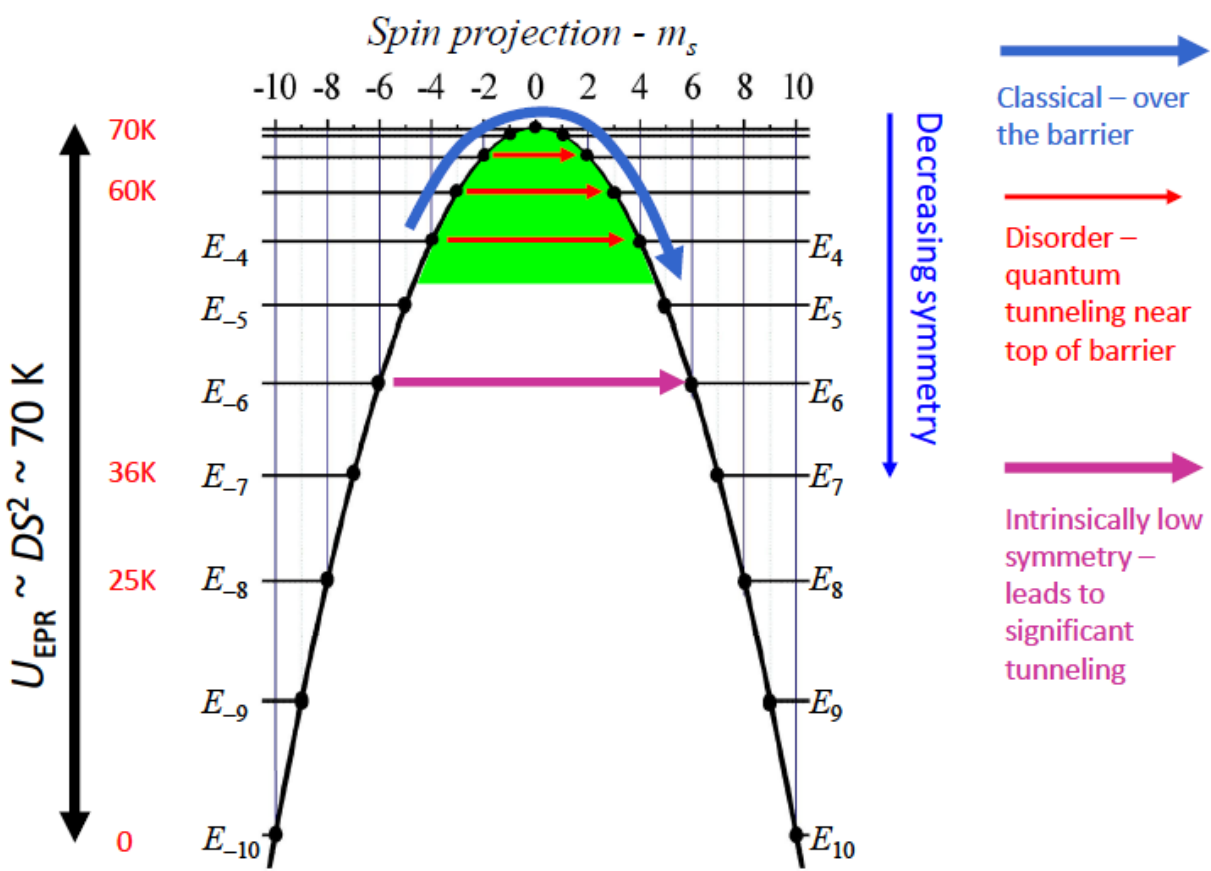

S. Hill et al., Figure 11

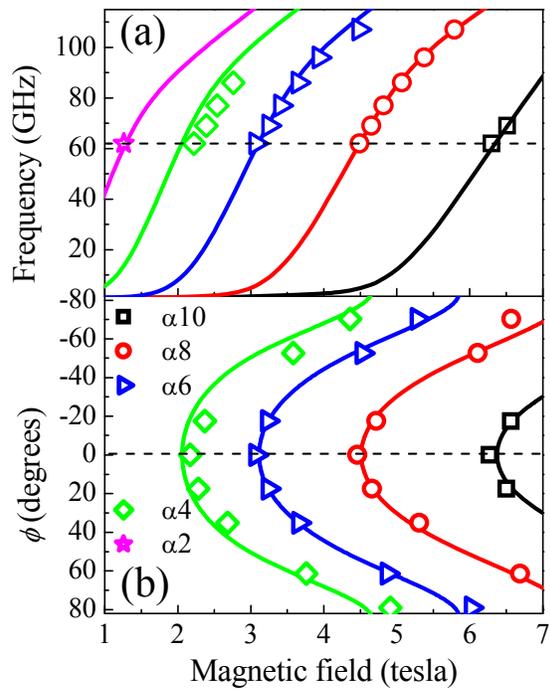

S. Hill et al., Figure 12

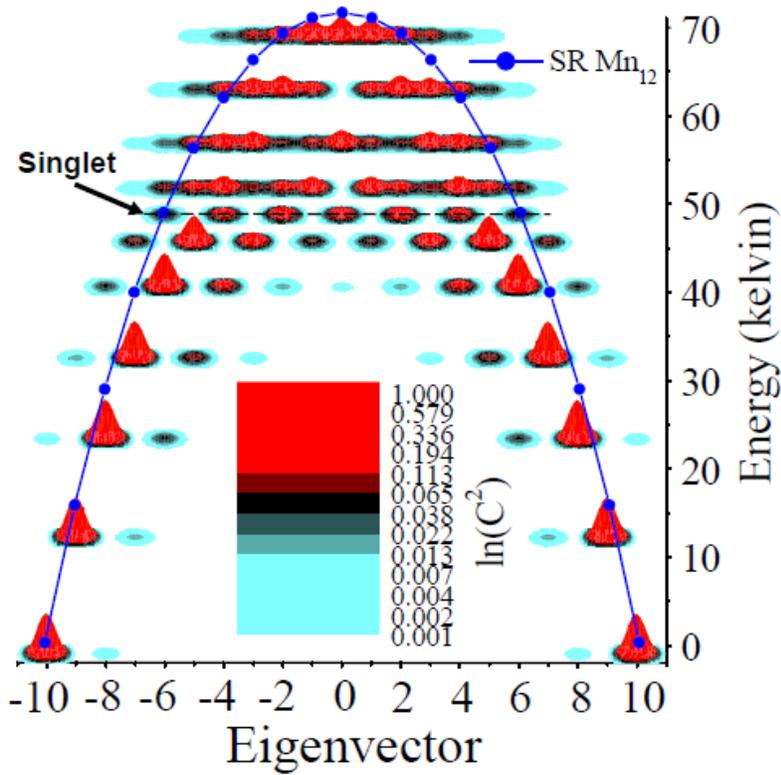

S. Hill et al., Figure 13